\documentclass[oldversion]{aa}
\usepackage{natbib}
\usepackage{txfonts}
\usepackage{times}
\usepackage{graphicx}
\def\v{\mathrm v}
\newcommand{\ds}{\displaystyle}
\begin{document}
\newcommand{\Rem}[1]{}
\title{Heating of the cooling flow}
\subtitle{The feedback effervescent heating model}
%\title{ Why is the cooling flow in clusters of galaxies a problem?}
%\subtitle{The end of cooling flow problem}
\author{Nasser Mohamed Ahmed \inst{1}}
%
%\offprints{E-mail:nasser@astro.rug.nl}
\institute{National Research Institute of Astronomy and Geophysics (NRIAG), El-Marsed Street, 11421 Helwan, Cairo, Egypt }
%
%\institute{Kapteyn Astronomical Institute, Netherland- Groningen \and National Research Institute of Astronomy and Geophysics (NRIAG), El-Marsed Street, 11421 Helwan, Cairo, Egypt }
\date{}
%\abstract{}{}{}{}{}
% 5 {} token are mandatory
\abstract
   {The standard cooling flow model has predicted a large amount of a cool gas in the clusters of galaxies. The failure of the Chandra and XXM-Newton telescopes to detect a cooling gas (below 1-2 keV) in the clusters of galaxies has suggested that some heating process must work to suppress the cooling. The most likely heating source is the heating by AGNs. There are many heating mechanisms, but we will adopt the effervescent heating model which is a result of the interaction of the bubbles inflated by AGN with the intra-cluster medium(ICM).

Using the FLASH code, we have carried out  1D- time dependent simulations to investigate the effect of the heating on the suppression of the cooling in cooling flow clusters. We have found that the effervescent heating model can not balance the radiative cooling and it is an artificial model. Furthermore, the effervescent heating  is  a function of the ICM pressure gradient but the cooling  is proportional  to the gas density square and square root of the gas temperature.} {}{}{}{}
\maketitle 
\section{Introduction}                               %==========================
According to the steady flow assumption of the standard cooling flow model, we must find a cool gas and a multi-phase medium within the cooling core in clusters of galaxies which are not observed in any wavebands  (X-ray and non X-ray). This is known as the cooling flow problem in clusters of galaxies. In others words, there is  a discrepancy between standard cooling flow model and observations (X-ray and non X-ray). This strong  discrepancy is  interpreted as either the gas is being prevented from cooling by some heating process, or  it cools without any spectroscopic signature~\citep{fabian01} which is difficult.  But, we see that this discrepancy between the standard cooling flow model and X-ray observations indicates that either  the cooling in the center of cooling flow clusters must be suppressed by any heating mechanism, or  the steady flow assumption of the standard cooling flow model is not appropriate. Somewhere else, we have concentrated in the second point which it is found that the steady flow with cooling is impossible, i.e the cooling flow problem is due to the wrong steady flow assumption. In this work we will concentrate in the first point, heating the cooling flow.\\
The failure of the multi-phase model has revived the idea of a heating mechanism which can suppress the cooling. There are five main conditions for the heating \citep[see][ for review]{gardini04}~: 
\begin{enumerate}
\item  The heating must be fine tuned and distributed to get the smooth  observed temperature profile. Too much heating would result a outflow from the center region. Too little heating is not sufficient to suppress the cooling. Moreover, the heating process must be self regulated: the mass flow rate triggers the heating and the heating reduces the mass flow rate~\citep{hans02}. That mechanism is called a heating with a feedback.
\item The heating mechanism by AGN must be sporadic because the radio activities are not observed in every cooling flow clusters.
\item  The kinetic energy injection must be subsonic and the ICM must not be shocked, as observed in general case. The shock could compress the gas producing the catastrophic cooling much faster.
\item  The heating mechanism must preserve the observed entropy profiles of the ICM,  which decrease toward the center.
\item  The heating must not destroy the metallicity profiles  which are peaked toward the centers \citep{tamura02,grandi01,irwin01}.
\end{enumerate}

A giant elliptical or cD galaxy sits at the potential well of every cooling flow cluster~\citep{mathews03,eilek04}. The popular heating mechanisms are a heating by AGN~\citep{binney95,binney93,br02}, thermal conduction, cosmic rays, galaxies motions, magnetic field reconnection \citep{soker90} and turbulent mixing \citep{kim03}. About 71 percent of the cDs in the cooling flow clusters are radio loud compared to only 23 percent of non-cooling flow clusters ~\citep{burns90}. This result suggests that there a relationship  between the AGN  activities and the presence of cooling flow.

In some of cooling flow  clusters, the recent X-ray observations reveal  holes in the X-rays surface brightness coincident with the radio lobes, known as X-ray cavities or bubbles~\citep[ see][ as example]{fabian06}. The  bubbles or cavities are not a universal phenomenon, suggesting a duty cycle. {}~Radio sources are not even detected in some cooling-flow clusters.

The observed metallicity gradients make a constraint on the ability of baubles to mix the ICM~\citep{bohringer04}. ~\citet{bm102} have run many 1D simulations of clusters with various levels of heating. They concluded that the best fit of temperature profiles with the real clusters is without heating. The AGNs are assumed to inject buoyant bubbles into the ICM, which heat the ambient medium by doing work $PdV$ as they rise and expand. This mechanism is called the  effervescent heating model or mechanism~\citep{begelman01}. In this work, we will concentrate on this mechanism, using time dependent hydrodynamics simulations, FLASH code~\citep{fry02}.
%
%=======================================================================
%=======================================================================
%=======================================================================
\Rem{
\section{The cooling time scale and the heating process}   %============
%=======================================================================
The cooling time $t$ is   the period of time for a given system to cool from a higher temperature to a lower one. The cooling time scale $t_{cool}$ is a approximation for it. The term {\it scale} is used to denote that approximation.  In this section, we will see the effect of heating on the cooling time scale. In the most   of the cooling flow clusters centers, the cooling time scale is in the range of $10^{8 }-10^{9}$ Gy less~than the age of cluster $t_H$ but there is only a cooling gas~ in the range of 1-2 keV. This  suggests that the heating process must work to suppress the cooling, but we will see that is  absurd.  The cooling time scale is given by~\citep{pet_fab06}:
\[ t_{cool} \;=\; a \frac{\sqrt T}{n} \] 
where {\it a} is constant.   $T$ is the gas temperature and $n$ is the gas particle number density.

We can suppose that there are  two clusters with the same initial condition , but the first one is only cooling and the second one is cooling and heated. The cooling time scale in the first one is given by:
\[ t_{cool-1} \;=\; a \frac{\sqrt T}{n} \;\; \approx \; t_H \] 
This cooling time scale should be near to the age of cluster $t_H$ if the cooling time scale is good estimator for cooling time. The cooling time scale in the second case is:
\[  t_{cool-2} \;=\; a \frac{\sqrt{T+ \Delta T} }{n- \Delta n} \]
and
\[ t_{cool-2}  \; > \;  t_{cool-1}\]
The heating process increases the gas temperature by  $\Delta T$ and decreases the gas density by $\Delta n$, by suppressing the flowing gas. The result is that the heating process increases the cooling time scale. If the heating process is taking place in the centers of cooling flow clusters, we must find that the cooling time scale is in the order of $10^{10}$ years or higher, which is not observed. This result means that the cooling time scale is not a good estimator for the cooling time. In the following sections, we will see the effect of heating on the cooling time scale using simulation.
}
%
%================================================================================================
%================================================================================================
%================================================================================================
%================================================================================================
%================================================================================================
%================================================================================================
\section{Effervescent heating model}                                        %%%%%%%%%%%%%%%%%%%%%
%================================================================================================
%
The center AGN is assumed to inflate bubbles of relativistic plasma in the ICM. These bubbles will expand as they rise, doing $PdV$ work on their surroundings. {\it Assuming a steady state and spherical symmetry,} the energy available for heating the ICM known as the effervescent heating~\citep{begelman01,ruszk02}~ is:
\[ \dot{E} \;\propto P_b(r)^{(\gamma_b -1)/\gamma_b} \; , \]
where $P_b$ is the partial pressure of buoyant fluid inside the bubbles at radius $r$,  and $\gamma_b$ is the adiabatic index of buoyant fluid. ~\citet{begelman01} and ~\citet{ruszk02} have assumed that the  pressure of a buoyant fluid
is scaled with the thermal pressure of the ICM; i.e.  the ratio between the pressure of a buoyant fluid and  the thermal pressure of the ICM is a constant with radius. In this case:
\[ \dot{E} \;\propto P(r)^{(\gamma_b -1)/\gamma_b}\]
In that model, we should note that the heating profile by AGN is proportion to the ICM properties, not the AGN itself. The heating rate per unit volume is given by:
\begin{equation}
\mathcal{H}_\v \; \approx \; - h(r) \; \nabla \cdot \frac{\dot{E}}{4\pi r^2} \;=\; - h(r)\; \left( \frac{P}{P_o} \right)^{(\gamma_b-1)/\gamma_b} \frac{1}{r}\; \frac{d\ln P}{d\ln r} \; , \label{eq:heat_model}
\end{equation}
where $\mathcal{H}_\v$ is heating rate ($erg\; cm^{-3}\;Sec.^{-1}$),~  $P_o$ is the center pressure and $h(r)$ is a normalization function multiplied by a cutoff function $(1-e^{-r/r_o} )$ as proposed by~\citet{ruszk02}~:
\[ h(r) = \frac{L}{4\pi r^2}\; \; q^{-1}\;\; \times \; (1-e^{-r/r_o} ) \; , \]
where $r_o$  is a small cutoff radius (10-25 kpc) and $L\;$  is the Luminosity of the center source which is given by:
\[ L \;=\; \int_o^{r_{max}} \mathcal{H}_\v \; dV \]
Upon reflection,   the function  $( 1-e^{-r/r_o} )$ is not a cutoff function as proposed, but it is  {\it a shallowness function}  added to let the heating profile  shallower more than the original one, working as a artificial fine tune.  The function $e^{-r/r_o}$ is working as a cutoff function. As  $r_o$ increases, we get a shallower profile. \\
The factor $q$ is given by
\[ q \;=\; - \int_{0}^{r_{max}} \; \left( \frac{P}{P_o} \right)^{(\gamma_b-1)/\gamma_b} \frac{1}{r}\; \frac{d\ln P}{d\ln r} \; (1-e^{-r/r_o} ) \; dr\]
We found that without the minus sing (not in the original paper)  heating will be negative. This factor $q$ is a  normalization factor to  let the total heating inside $r_{max}$ be equal to $L$.~~ By doing this normalization, we find that the presence of the term $P_o$ (in equation~\ref{eq:heat_model}) is not essential. \\
Finally, the luminosity of the central source $L$ is given by  a feedback from mass flow rate (mass accretion rate)
\[ L \;=\; \epsilon \dot{M}_{flow-min} \;\; c^2 \]
where $\epsilon$  is the accretion efficiency and c is the  speed of light.  $ \dot{M}_{flow-min}$ which is given by:
\[ \dot{M}_{flow-min}~=(-4\pi r^2_{min}\;\rho \v) \]
is the mass flow rate in the inner radius $r_{min}$ depending on the resolution of simulation. For high resolution simulation, we are aware that the mass flow rate is a small quantity which is not enough to fuel the AGN but that model is most accurate one we know. The actual heating mechanism is still unclear and uncertain.\\ \\
At the end, we get the heating profile as:
\begin{equation}
\mathcal{H}_\v \;=\; - \frac{L}{4\; \pi\; q\; r^2}\; P^{\frac{-1}{\gamma_b}}\; \frac{dP}{dr}\; (1-e^{-r/r_o} ) \label{eq:heat_profile}
\end{equation}
and the internal energy equation becomes:
\begin{equation}
\frac{\partial H_\v}{\partial t} +\nabla(H_\v\; \vec{\v}) -\frac{dP}{dt} \;=\; -\epsilon_\v +  \mathcal{H}_\v
\end{equation}
where $\epsilon_\v$ is the X-ray emissivity and $H_\v$  is the gas enthalpy per unit volume $(U_v+P)$, and $H_\v$  is given by:
\[ H_\v\;=\; \frac{5}{2}\frac{k\;T\;\rho_g}{\mu\; m_p}\]
where $\rho_g$ is the gas density.
%
%
%
%========================================
\subsection{Problem with the effervescent heating model}
%========================================
The effervescent heating profile depends on the gradient of ICM pressure rather than temperature and it is much  steeper than the X-ray emissivity. This model needs  a fine tuning because the velocity in the center must be a negative quantity, otherwise the gas could be accumulated somewhere near to the center. The gas will flow from outwards and inwards into that region. In other words, the feedback can not ensure the fine tuning and can cause the velocity in the center to be a positive quantity and in the other part of the cooling core to be negative. \\ \\
\citet{ruszk02} claimed that their model does not need a fine tuned heating. We argue that  their model is fine tuned by choosing some value of $r_o$ to control the steepness of the heating profile, and choosing $\epsilon$ accretion efficiency to control the amplitude of the heating function.  The accretion efficiency is not a constant for any simulation but is allowed to change to guarantee an artificial fine tuned heating. \\
The scaling of  the  buoyant fluid pressure inside the bubbles with the ICM pressure is a difficult assumption, during the motion of bubbles in ICM. The bubbles must expand in some way to keep this ratio a constant. The free parameters ($r_o$, $r_{max}$,  $\epsilon$, and  $r_{min}$) in the effervescent heating model are responsible for the model to be artificial and do not reflect any  physics for the heating.
%
%==========================================================================================
%==========================================================================================
\section{Initial conditions}
%==========================================================================================
%==========================================================================================
%% %%
The total cluster mass or the gravitation acceleration can be determined from X-ray observations by assuming that the gas is in a hydrostatic equilibrium and a spherical symmetry
\begin{equation}
\vec{g}\;=\; \frac{\nabla P}{\rho_g} \label{eq:hydro_static_xray}
\end{equation}
where $P$ and $\rho_g$ are the observed gas pressure and density. The inflowing gas has  nearly no effect on the total cluster mass, then we can assume that the value of $ \vec{g}$ nearly does not change since the cluster started to cool.\\

In cooling flow clusters, the temperature and gas density profiles change only in the cooling core region due to cooling, but they do not change in the  outer part of the cooling core. The cooling has no effect on the total clusters masses.   We determine  the initial temperature profile by fitting the  observed X-ray temperature profile of the  outer part of the cooling core with the universal temperature profile of~\citet{loken02}. \\
\begin{equation}
\displaystyle
\frac{T(r)}{T_x}=T_o \; (1+a_x \; r/r_{vir})^{-\delta}
\end{equation}
where $T_o$, $a_x$, and $\delta$ are free parameters. For most clusters, these fitting parameters are found to be $a_x = 1,5$ , $\delta=1.67$, and $T_o = 1.31$~.
Then the gas density is given by
\begin{equation}
\rho_g(r)\quad=\quad  \rho_{go} \; \times \quad \left( \quad \frac{T_{go}}{T_g(r)}\;\; \times \quad   \exp^{\; \ds \; \frac{\mu m_p}{k} \int_{\;0}^{\;r} \; {\frac{g(r)}{T_g(r)} \;dr }  } \quad \right)
\end{equation}
where $\rho_g$ is the gas density.  $\rho_{go}$ and $T_o$ are the gas density and temperature at the center. The $g(r)$ is obtained by current X-ray observation assuming a hydrostatic equilibrium as in equation~\ref{eq:hydro_static_xray}.

\begin{figure}
\resizebox{\hsize}{!}{\includegraphics[angle=90]{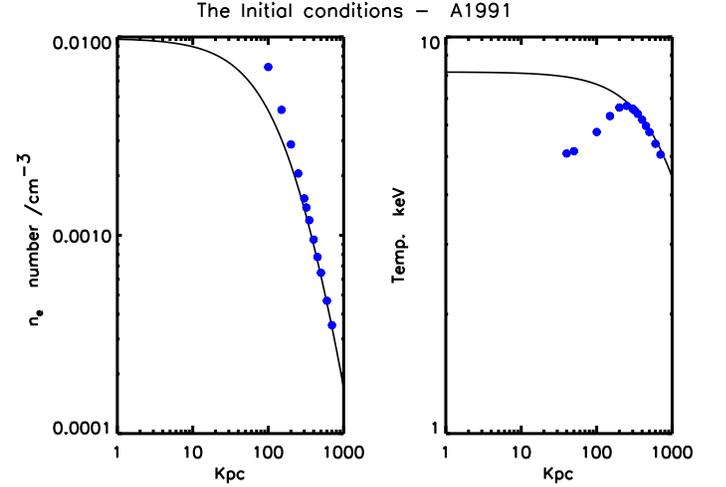}}
\caption{Initial conditions: the dots are the observed X-ray temperature and gas density profiles \citep{Vikhlinin05}. The solid line is the initial fitted temperature profile \citep{loken02}.}
\label{fig:a1991_init_cond}
\end{figure}

We will take the observed X-ray gas and temperature profiles of the Chandra observation~\citep{Vikhlinin05} as an initial conditions for the cluster A1991 as in figure~\ref{fig:a1991_init_cond}.  %
%
%
%=======================================================================
\subsection{Models}                              %======================
%=======================================================================
%
We have carried out four model simulations assuming the effervescent heating model. Model A is carried out only with   cooling. Model B is a model with an effervescent heating and cooling and model C is the same as model B but with higher resolution (smaller value of $r_{min}$). Finally, the model D has a  higher cutoff radius $r_o$ to get a smoother heating profile, see table~\ref{tab:models_heat}.
\begin{center} 
\begin{table*}[htbp]
\caption{The models}
\label{tab:models_heat}
\begin{tabular}{|c||c|c|c|c|c|}
\hline
Model    & Note                & cutoff radius $r_o$ &  $ r_{max} $ &   $\epsilon$ & Resolution -$r_{min}$ \\
\hline \hline
Model  A & Pure Cooling        & -                   &     -        &      -       & 520 pc     \\
\hline  Model  B & Cooling and Heating & 22 Kpc              &  2000 Kpc    &  .01         & 520 pc     \\
\hline    Model  C & Cooling and Heating & 22 Kpc              &  2000 Kpc    &  .01         & 78 pc     \\
\hline
Model  D & Cooling and Heating & 40 Kpc              &  2000 Kpc    &  .001        & 520 pc     \\
\hline
\end{tabular} 
\end{table*}
\end{center}
%%
%===================================================================================================
%===================================================================================================
\section{Results and discussion}
In this section, we describe  our four simulations and the results will be given as follow:
\subsection{Model A}
\begin{figure}
\resizebox{\hsize}{!}{\includegraphics[angle=90]{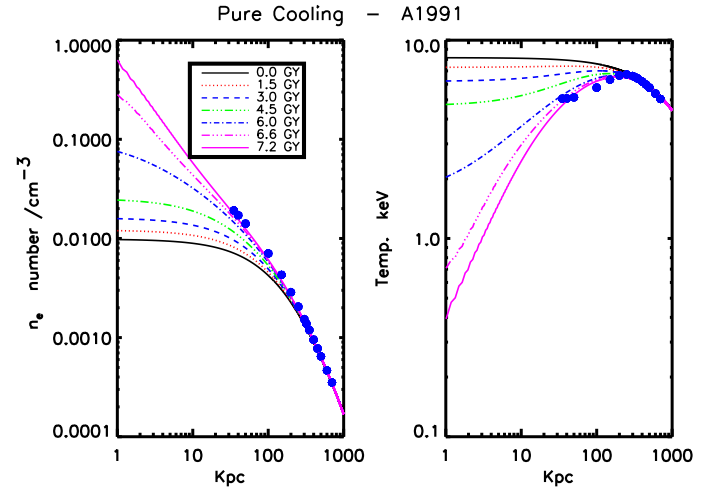}}
\caption{Model A, pure cooling: ~the simulated gas density and temperature profiles at different times up to  7.2 G yr. The big dots represent the observations. The simulation agrees very will the X-ray observation~\citep{Vikhlinin05}.}
\label{fig:a1991_cooling}
\end{figure}

The cluster A1991 is observed by the Chandra X-ray observatory~\citep{Vikhlinin05}. Model A is a time dependent pure cooling simulation, there is no heating. This simulation have run until time 7.2 Gy where the simulated temperatures and densities agree very well with the X-ray observations~\citep{Vikhlinin05} (see figure~\ref{fig:a1991_cooling}).
After time 7.2 Gy, the catastrophic  cooling accrued only within a few cells due to the resolution, reflecting boundary conditions and steep potential profile.  The heating can not help because it will heat the whole cooling core.
%
%-------------------------------------------------------------------------------------------------
\subsection{Model B}
\begin{figure}
\resizebox{\hsize}{!}{\includegraphics[angle=90]{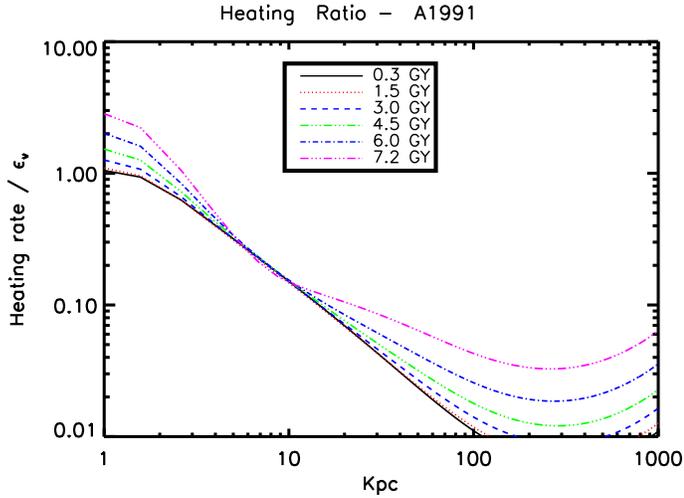}}
\caption{Model B:the plot of the ratio between the heating rate and cooling rate~~($\mathcal{H}_\v \;/\; \epsilon_\v $). }
\label{fig:a1991_heating_ratio}
\end{figure}

\begin{figure}
\resizebox{\hsize}{!}{\includegraphics[,angle=90]{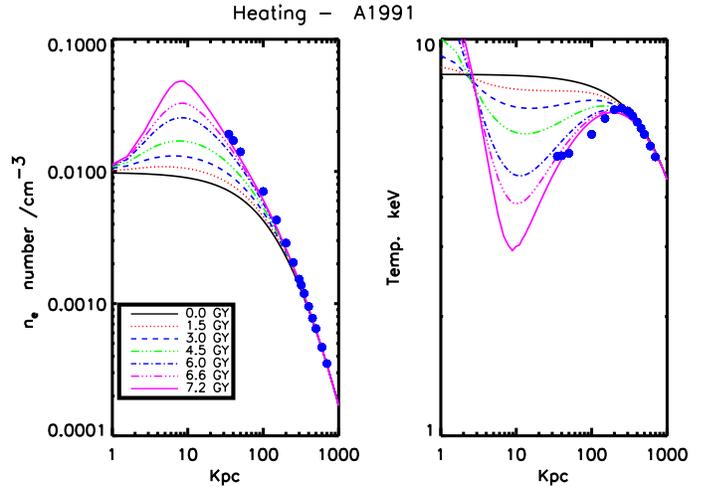}}
\caption{Model B:the simulated gas density and temperature profiles as functions of radius. The big dots represent the observations}
\label{fig:a1991_heating}
\end{figure}

%------------------
The  mass accretion rate (to fuel the black hole)  is taken to be equal to the mass inflow rate at the inner radius $r_{min}$.  In this model, we have set the cutoff radius $r_o$ and heating efficiency $\epsilon$ at $22$ kpc and $.01$ respectively,~ see table~\ref{tab:models_heat}. The resolution in the center is about $520$ pc which is large than the black hole accretion radius. \citet{ruszk02} have chosen $r_{min}$ equal to $1$ Kpc to get large mass flow rate which can fuel the black hole in the cluster center. We are certain that at small radius the mass flow rate is not sufficient to fuel the central black hole to stop the cooling. However, that is the only way to study the effect of the heating in clusters properties. The heating profile is much  steeper than the X-ray emissivity which makes it difficult  for the heating profile  to balance the radiative cooling. In figure~\ref {fig:a1991_heating_ratio}, we plot the heating rate and cooling rate ratio.
As in figure~\ref{fig:a1991_heating},  the cluster center is overheated and there is no balance between the heating and cooling; that is clear in the temperature profile.  Moreover, inside $10$ kpc, the gas density peaks near to center not in the center which is not observed. The gas  becomes a clumpy structure. There is a outflow from the cluster center.   %
%
%
%
%--------------------------------------------------------------------------------------------------
%==============================
\subsection{Model C}
%==============================
In model B, the inner radius (resolution) is about 520 pc which is larger than the black hole accretion radius. This model is similar to  model B but the resolution is higher to get  smaller accretion radius (inner radius). We have set the resolution or inner radius equal to $78$ pc which can not be  achieved by three dimensions simulations. The result is that the heating is not enough to balance the radiative cooling because the mass flow rate at a small radius is very small . Of course this model is  more realistic than model B because of the smaller inner  radius (accretion radius). \\
\begin{figure}[htbp]
\resizebox{\hsize}{!}{\includegraphics[angle=90]{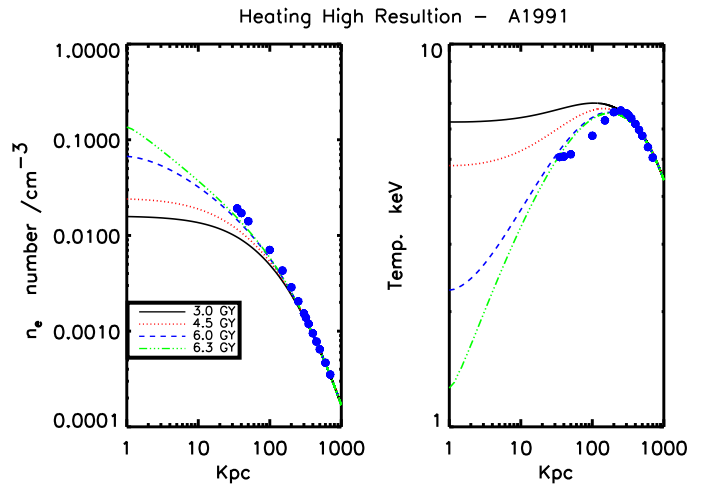}}
\caption{Model C, high resolution (78 pc)and heating with cooling: the simulated gas density and temperature profiles are the same as in pure cooling simulation. The big dots represent the observations}
\label{fig:a1991_heating_high}
\end{figure}
%% %%
%======================================================================================================
%==============================
\subsection{Model D}
\begin{figure}
\resizebox{\hsize}{!}{\includegraphics[angle=90]{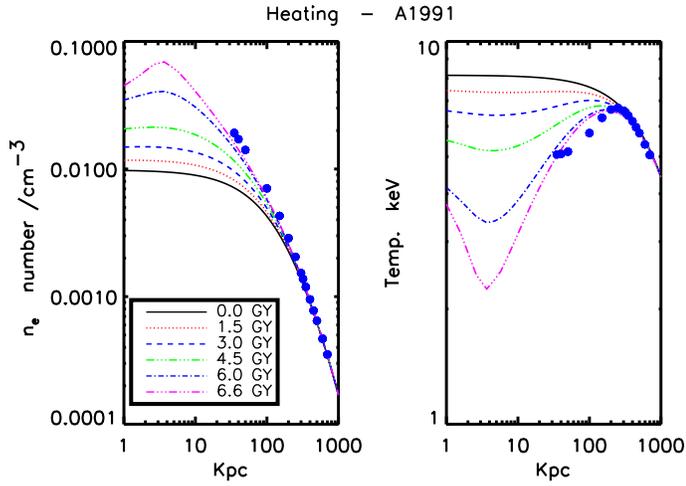}}
\caption{Model D, cooling with heating: the simulated gas density and temperature profiles as  functions of radius. The big dots represent the observations}
\label{fig:a1991_heating1}
\end{figure}

%==============================
In the model D, we set cutoff radius ($r_o=44$ kpc) at a larger value than in model B in order to get a smoother heating profile, see table~\ref{tab:models_heat}. The result is much better than model B but there is still a problem to fit the observations, see figure~\ref{fig:a1991_heating1}. The temperature at the center is   higher than observed. ~With the large cutoff radius $r_o$, the heating profile can balance the cooling but it becomes an artificial model which is not physical. In other words, the  gradient of the heating profile becomes near to the gradient of the X-ray emissivity, ensuring the balance between the heating and cooling.   %
\begin{figure}
\resizebox{\hsize}{!}{\includegraphics[angle=90]{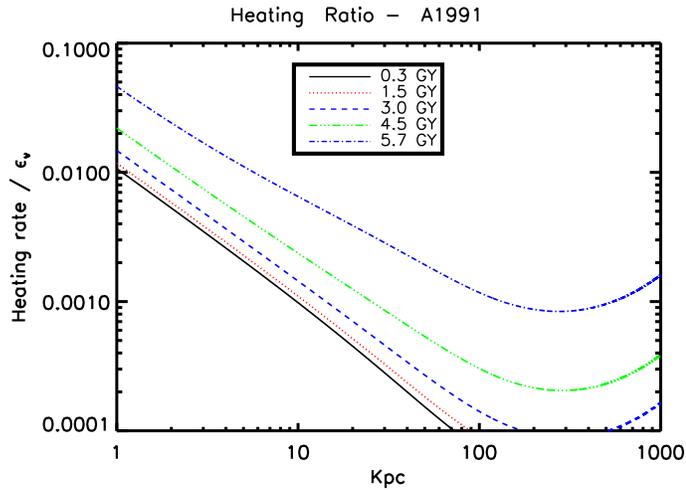}}
\caption{Model D ($r_o=44$ kpc): the plot of the ratio between the heating rate and cooling rate($\mathcal{H}_\v \;/\; \epsilon_\v$). With increasing the the cutoff radius $r_o$, the heating profile becomes more smoother.}
\label{fig:a1991_heating_ratio1}
\end{figure}
%
%
%===========================================
\subsection{The effect of the heating on cooling time scale}
\begin{figure}
\resizebox{\hsize}{!}{\includegraphics[angle=90]{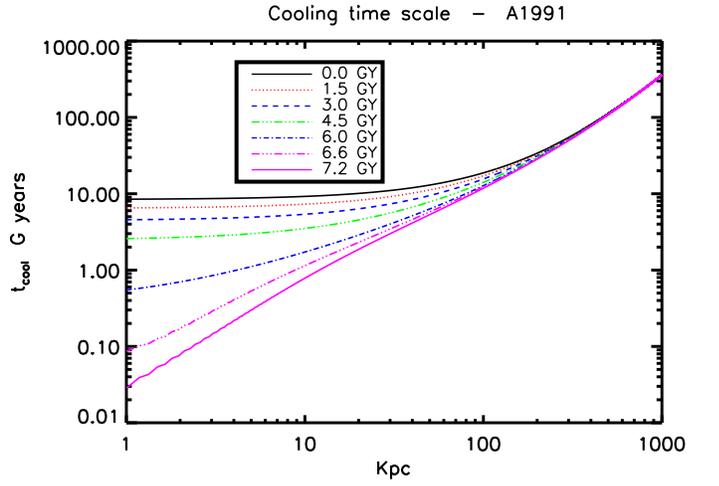}}
\caption{Model A, pure cooling: the cooling time scale is very short in the center, even  there is no much cooling.}
\label{fig:a1991_purecooling_tmscl}
\end{figure}

\begin{figure}[htbp]
\resizebox{\hsize}{!}{\includegraphics[angle=90]{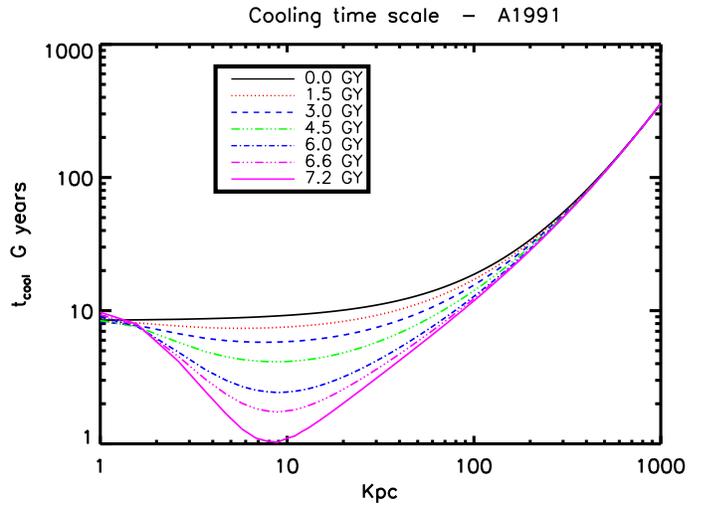}}
\caption{Model B: the cooling time scale is higher than pure cooling simulation. In the center, the cooling time scale is higher than observed which is against the heating of cluster.}
\label{fig:a1991_heating_tmscl}
\end{figure}

%============================================
In some cooling flow clusters, the observed cooling time scale is very short, in the range of $10^{8}$ yr to $10^{9}$ yr, but the cooling gas below 1-2 keV is not observed. This result is interpreted  as indication that there must be a heating mechanism to stop the cooling. In figure~\ref{fig:a1991_purecooling_tmscl}, The cooling time scale is very short in the center, even there is no catastrophic cooling or there is no much cooling. Figure~\ref{fig:a1991_heating_tmscl} show that  the heating increases the cooling time scale. If the cluster was heated, then we should find that the cooling time scale must be  higher than observed.   %
%530 Pc  78.487623Pc  file:///home/nasser/A1991/a1991_purecooling_tmscl.eps a1991_heating_ratio1.eps
%
%%
%%
%=================================================================================
%********************************************************************************* 
\section{Summary and conclusion}  \label{sec:sm_heat_model}                   %%==
%=================================================================================
%********************************************************************************* 
We have carried out time dependent simulations with cooling and heating. The general result is that the best fit to the observations is the model without heating (model A). The effervescent heating  is  a function of the ICM pressure gradient but the cooling  is proportional to the gas density square and square root of the gas temperature.~Furthermore, the conclusions are~: \\ \\
1- From our simulations, the cooling can not be a steady flow as assumed by standard cooling flow model. The gas density must increase with the time; i.e the ICM must be compressed under the force of the inflowing gas.\\ \\
2- The effervescent heating model profile (without the smoothness function) is more steeper than the radiative cooling profile, which makes it is very difficult to balance the cooling. \\ \\
3-  The function $1-e^{-r_o/r}$ is not a cutoff function but it is a smoothing function in order to control the steepness of the heating function letting it ~balance the radiative cooling; i.e. it is an artificial model.  \\ \\
4- At inner radius close to the realistic value of  the accretion radius (model D), the accretion mass rate due to flowing gas is not enough to fuel the black hole in the center of cluster. It's not reasonable to set the accretion radius of the black hole at large radius, (for example 1-1.5 kpc, as used for the three dimension simulations). \\ \\
5- In some cooling flow clusters, the observed cooling time scale is very short, in the range $10^{8}$ yr to $10^{9}$ yr, but the cooling gas below 1-2 keV is not observed. This result is interpreted  as indication that there must be a heating mechanism to stop the cooling. As in figure~\ref{fig:a1991_purecooling_tmscl}, the cooling time scale inside a radius of 10 kpc is shorter than $1$ Gy but there is no a cool gas.
Moreover, from our simulations (model B),  we found that the heating process increases the cooling time scale. If cooling time scale is a good approximation for  the actual cooling time and the cluster is heated, then we must find that the cooling time scale is larger than observed.\\ \\
\begin{acknowledgements}
-  I have used FLASH code for my hydrodynamics simulations. The FLASH code was in part developed by DOE-supported ASC/Alliance Center for Astrophysical Thermonuclear Flashes at the University of Chicago. I acknowledge their work for that code which can handle very difficult problems in astrophysics with an adaptive mesh technique.
\end{acknowledgements} 
\bibliographystyle{aa}

\end{document}